\documentclass{aa}
\usepackage{txfonts}
\usepackage{natbib}
\usepackage{graphicx}

\bibpunct[, ]{(}{)}{;}{a}{}{,}

\def\llm{{\sc LLmodels}}

\def\detail{{\sc DETAIL}}

\def\vald{{\sc VALD}}
\def\logg{\log g}
\def\teff{T_{\rm eff}}

\def\taunu{\log\tau_{\rm 5000}}

\def\bs{$\langle B \rangle$}
\def\paper1{Paper~I}
\def\halpha{H$\alpha$}

\def\ddafit{{\sc DDAFIT}}
\def\synthmag{{\sc SYNTHMAG}}
\def\ei{$E_{\rm i}$}
\def\loggf{$\log(gf)$}
\def\hd{HD~24712}

\begin{document}

\title{Model atmospheres of chemically peculiar stars}
\subtitle{Self-consistent empirical stratified model of \object{\hd}}

\author{D. Shulyak\inst{1} \and T. Ryabchikova\inst{1,2}  \and L. Mashonkina\inst{2} \and O. Kochukhov\inst{3}}
\offprints{D. Shulyak, \\
\email{denis@jan.astro.univie.ac.at}}
\institute{Institute of Astronomy, Vienna University, Turkenschanzstrasse 17, 1180 Vienna, Austria \and
Institute of Astronomy, Russian Academy of Science, Pyatnitskaya 48, 119017 Moscow, Russia \and
Department of Astronomy and Space Physics, Uppsala University, Box 515, 751 20, Uppsala, Sweden}

\date{Received / Accepted}

\abstract
{High-resolution spectra of some chemically peculiar stars clearly demonstrate the presence
of strong abundance gradients in their atmospheres. However, these inhomogeneities are usually ignored
in the standard scheme of model atmosphere calculations, braking the consistency between 
model structure and spectroscopically derived abundance pattern.}
{In this paper we present first empirical self-consistent stellar atmosphere model of roAp star \hd\, 
with stratification of chemical elements included, and which is derived
directly from the observed profiles of spectral lines without time-consuming simulations 
of physical mechanisms responsible for these anomalies.}
{We used the \llm\, stellar model atmosphere code and \ddafit\, minimization tool for analysis of
chemical elements stratification and construction of self-consistent atmospheric model. 
Empirical determination of Pr and Nd stratification in the atmosphere of \hd\, is based on NLTE line formation 
for \ion{Pr}{ii/iii}\, and \ion{Nd}{ii/iii} with the use of the \detail\ code.}
{Based on iterative procedure of stratification analysis and subsequent re-calculation of
model atmosphere structure we constructed a self-consistent model of \hd, i.e. the model
which temperature-pressure structure is consistent with results of stratification analysis. It is shown that
stratification of chemical elements leads to the considerable changes in model structure as to compare with
non-stratified homogeneous case. We find that accumulation of REE elements allows for the inverse temperature
gradient to be present in upper atmosphere of the star with the maximum temperature increase of about $600$\,K.}


\maketitle

\section{Introduction}
Atmospheres of chemically peculiar (CP) stars exhibit the presence of inhomogeneous distribution of chemical elements.
These inhomogeneities are frequently seen as abundance spots on stellar surfaces \citep[see, for example,][]{di1,di2,di3} 
and as clouds of
chemical elements concentrated at certain depths along the stellar radii \citep{str1,str2,str3}. 
Several tools have been developed
and successfully applied to the analysis of element distributions based on detailed analysis
of observed line profiles. The horizontal elements distribution is investigated using the rotational modulation of
spectral lines: as star rotates, the abundance spots on its surface should reveal themselves 
as variation of line profiles of respective elements observed at different rotational phases of the star. 
The automatic procedure of finding the spots position and abundance
values via the comparison of observed and calculated spectra is performed via Doppler Imaging techniques firstly introduced by
\citet{goncharsky} and further developed by \citet{piskunov_wehlau} (see also \citet{di2}).
The vertical distribution of chemical elements is modeled assuming the stratification profiles of chemical elements 
to be represented by simple step-function with two different abundances in upper and lower atmosphere as was suggested
in \citet{babel}, or, as recently showed by \citet{vip}, applying the regularized minimization algorithm.
The results of these analyses show that strong abundance gradients do exist in atmospheres of CP stars 
with accumulation of various chemical elements at different atmospheric layers.

From theoretical point of view, the appearance of inhomogeneous element distribution with high abundance 
gradients in atmospheres of CP stars is explained by microscopic radiative diffusion processes: the
atmospheres of these stars are believed to be dynamically stable enough 
for radiative levitation and gravitational settling to rule the distribution of chemical elements \citep{michaud}.

These abundance inhomogeneities may have rather strong influence on model atmosphere structure
via modification of opacity and emissivity coefficients and thus the modeling of element distributions 
in stellar atmospheres
should, in principle, include the simulation of these diffusion processes on every step of model atmosphere calculation.
Such models were successfully developed and improved in last years \citep[see, for example,][]{diff1,diff2},
however, the sensitivity of modeled element distributions to the input physics (that is frequently poorly understood), 
the absence of accurate atomic data for some elements, and high computational expenses 
do not allow implementation of these models 
for detailed modeling of theoretical line profiles and their comparison with high-resolution observations: 
although the theoretical stratification is able to predict some observed characteristics of CP stars, 
it is still not applied for the quantitative interpretations of modern high-resolution
spectroscopic observations.

One of the necessary ingredient in analysis of elements stratification is the computation of synthetic spectra 
and their subsequent comparison with observations: a set of such computations
is needed to construct the stratification profile of a given element that ensures a best fit 
between theoretical and observed line profiles. 
These empirical investigations of elements stratification in stellar atmospheres are very important 
for understanding the physical
conditions in atmospheres of CP stars and, ideally, should provide the test ground for recent theoretical calculations.
The only methodological difficulty in empirical analysis is that almost in all cases 
the synthetic spectrum is computed using standard model atmospheres,
i.e. model atmosphere that were computed under the assumption of homogeneous elements distribution.

To avoid this inconsistency we made an attempt to implement an iterative procedure 
of vertical stratification analysis with the subsequent re-calculation
of model atmosphere structure. This procedure is applied to cool roAp star \hd\, for which the stratification of 
Si, Ca, Cr, Fe, Sr, Ba, Pr, and Nd is derived.
Stratification analysis of Pr and Nd is based on non-local thermodynamic equilibrium (NLTE) line formation for \ion{Pr}{ii/iii}\, and 
\ion{Nd}{ii/iii}.
Using this approach we examine
how the stratification profiles of different elements are changing comparing with the case when 
only non-stratified model is applied and what is the overall effect of inhomogeneous elements distribution 
on model atmosphere structure.

In the next section we give an overview of observations used. Then, in Sect.~\ref{smethods}
we describe the general methods and atomic line data used
for analysis of element stratification. In Sect.~\ref{sres} we present the results
of stratification analysis and its effect on model atmosphere structure and some
observed parameters of the star. The main conclusions are summarized in Sect.~\ref{ciao}

\section{Observations}\label{sobs}
We used an average spectrum of \hd\, obtained at November~10/11, 2004, with HARPS (High Accuracy Radial velocity Planet Searcher) spectrometer 
at the 3.6-m telescope at ESO, La Silla. The details of the reduction procedure are given in \citet{puls}. 
In total 92 spectra were taken with 60\,s exposure time, signal to noise ratio S/N\,=\,120, and R\,=\,120\,000 resolving power. 
These spectra cover a spectral region from $3850$\AA\ to $6730$\AA. After averaging the final S/N\,=\,1000 was reached.  
HARPS spectra were obtained at the phase 0.867 close to magnetic maximum \citep{puls}. 
Photometric color-indices were taken from the SIMBAD electronic database\footnote{{\tt
http://simbad.u-strasbg.fr/simbad/}} with the additional Str\"omgren photometry from
\citet{martinez}, peculiar index $a$ from \citet{da1998}, and Geneva colors from \citet{burki}. 
We also used UV spectroscopic observations 
extracted from IUE data archive\footnote{{\tt http://ines.ts.astro.it/}}.

\section{Methods}\label{smethods}
\subsection{Stratification analysis}
To analyse the stratification of chemical elements in the atmosphere of \hd\ 
we applied step-function approximation as performed in \ddafit~--~an automatic
procedure for determination of vertical abundance gradients \citep{synthmag07} that was successfully
used in a number of studies \citep[e.g.][]{str1,str2,str4}.
In this routine, the vertical abundance distribution of an element
is described by four parameters: chemical abundance in the upper
atmosphere, abundance in deep layers, the position of the abundance jump
and the width of the transition region where chemical abundance changes between
the two values. All four parameters can be optimized simultaneously with the
least-squares fitting procedure and based on observations of unlimited number
of spectral regions. \ddafit\, also allows the stratification analysis with the magnetically-splitted lines.
For \hd\, we assumed a pure radial magnetic field with the modulus \bs=3.1\,kG \citep{puls}.
The spectral synthesis was performed using \synthmag\, code \citep{synthmag07}  which represents an improved version of the program
developed by \citet{P99}.

For Pr and Nd, we performed NLTE stratification analysis as 
described in \citet{pr-NLTE} and \citet{nd-NLTE} using a
trial-and-error method and the observed equivalent widths of the lines of the first and second ions.

\subsection{NLTE calculations}

The present investigation of \ion{Pr}{ii/iii}\, and \ion{Nd}{ii/iii}\, is based on the NLTE methods treated in our earlier studies 
\citep{pr-NLTE} and \citep{nd-NLTE}, where atomic data and the problems of line formation have been considered in detail. To describe 
briefly, comprehensive model atoms of praseodymium and neodymium include the measured and the predicted energy levels, in total, 6708 
levels of \ion{Pr}{ii}\, and \ion{Pr}{iii} and 2258 levels of \ion{Nd}{ii}\, and \ion{Nd}{iii}. The coupled radiative transfer and 
statistical equilibrium equations were solved using a revised version of the DETAIL program \citep{detail} based 
on the accelerated lambda iteration, which follows the efficient method
described by \citet{rh91,rh92}. 

As shown for Nd by \citet{nd-NLTE} and for Pr by \citet{pr-NLTE}, the main
non-LTE effect for their first ions in the model atmosphere representing the atmosphere of \hd, is overionization caused by a
super-thermal radiation of non-local origin close to the thresholds of the low-excitation levels. The departures from 
LTE for the lines of the first and the second ions are of the opposite sign and they are significant if the element is concentrated 
in the uppermost atmospheric 
layers where collisions are inefficient in establishing thermodynamic equilibrium. For the Pr and Nd distribution in the atmosphere 
of \hd\, determined by \citet{pr-NLTE} and \citet{nd-NLTE}, the NLTE abundance corrections, 
$\Delta\epsilon_{\rm NLTE}= \epsilon_{NLTE}-\epsilon_{LTE}$,
are positive for the lines of  \ion{Pr}{ii}\, and \ion{Nd}{ii}\, and constitute between $1.0$~dex and $1.4$~dex, 
while $\Delta\epsilon_{\rm NLTE}$ is 
negative for the lines of  \ion{Pr}{iii}\, and \ion{Nd}{iii}\, and ranges between $-0.3$~dex and $-0.7$~dex.

We performed NLTE calculations for \ion{H}{i} using the method
described in \citet{MZG07}. The model atom includes
levels with principal quantum numbers up to $n=19$. For the model
atmospheres investigated in this study, the ground state keeps their
thermodynamic equilibrium level populations throughout the atmosphere. In
the layers where the core-to-wing transition is formed, namely, between
$\log \tau_{5000} \simeq -1.2$ and $\log \tau_{5000} \simeq -2.2$, the $n =
2$ level is closely coupled to the ground state, while departures from LTE
for the $n = 3$ level are controlled by \halpha\, which serves as the
pumping transition resulting in an overpopulation of the upper level. For
\halpha, NLTE leads to weakening of the core-to-wing transition compared
to the LTE case.

\subsection{Model atmospheres}

To perform the model atmosphere calculations
we used the recent version of the \llm\, \citep{llm} stellar model atmosphere code (version 8.6). 
\llm\, is 1-D, hydrostatic, plain-parallel LTE code
which accounts for the effects of individual and stratified abundances. 
It is to note that the stratification of chemical elements is
an input parameter for \llm\, code and thus is not changing during model atmosphere calculation process. 
This allows to explore the changes in model structure due to stratification that was extracted directly from observations 
without modeling the processes that could be responsible for the observed inhomogeneities.

The following calculation settings were used for every model atmosphere calculation: 
the atmosphere of a star is discretized into $120$ 
layers between $\log\tau_{\rm 5000}=-8$ and $2$ with a non-equidistant spacing, 
i.e. with the higher points density in the regions of steep
abundance gradients to ensure accurate integration of radiation field properties and solution of other model equations. 
The frequency-dependent quantities are calculated with variable wavelength step: $0.1$ between $500$\AA\, and $20\,000$\AA, 
and $0.5$\AA\, between $20\,000$\AA\, and $70\,000$\AA\, respectively. 
\vald\ database \citep{vald1,vald2} was used as a source of atomic lines data 
for computation of line absorption coefficient.

The use of the LTE assumption in atmosphere modeling results from the following consideration. 
The opacity in the atmosphere of \hd\, is mostly contributed by neutral hydrogen. 
Our NLTE calculations for \ion{H}{i} showed no significant deviations from 
thermodynamic equilibrium populations for the $n=1$ level throughout the atmosphere 
and for the $n=2$ level below $\log \tau_{\rm 5000} = -3$. The departures from LTE are found 
to be larger for the hotter atmosphere of Vega, however, self-consistent NLTE modelling of 
\citet{nlte_vega} leads to only $200$~--~$300$~K changes in $T$ distribution compared 
to the LTE model in the layers between $\log \tau_{\rm 5000} = -1$ and $\log \tau_{\rm 5000} = -6$. 
The NLTE effect on atmospheric structure of \hd\, is expected to be smaller. Another important 
opacity contributors are the iron group elements. The iron and chromium among them are concentrated 
in deep atmospheric layers of \hd, below $\log \tau_{\rm 5000} = -2$ (see Fig.~\ref{Fstrat}), 
where the departures from LTE are expected to be small. This is confirmed by our NLTE calculations 
for Ca which distribution in the atmosphere is similar to that for the iron group elements (Fig.~\ref{Fstrat}).
We used the NLTE method for \ion{Ca}{i/ii} described by \citet{mash_ca}. 
We do not know the distribution of light elements, C, N, and O, however, according to \citet{ind} these elements 
do not introduce noticeable changes in model structure and emergent flux. 
The influence of the departures from LTE for two rare-earth elements (REE), Pr and Nd, 
on atmospheric structure of \hd\, is simulated in present paper.

Finally, the magnetic field with 
the intensity less than $5$\,kG  has marginal impact on model temperature-pressure structure
\citep{zeeman_paper1,zeeman_paper2}. 
Thus, we ignored the effects of polarized radiative transfer and Zeeman splitting in all 
model atmosphere calculations presented in this study.

\subsection{Atomic line data for stratification analysis}\label{atomic}
As was emphasized in \citet{Ryabchikova2003} the vertical abundance stratification manifests itself as an impossibility 
to fit the core and wings of strong lines with developed Stark wings 
(\ion{Ca}{ii}~K, \ion{Si}{ii}, \ion{Mg}{ii} lines, etc.)
with the same abundance or as an impossibility to describe the low and high-excitation lines of the same ion
with a chemically homogeneous atmosphere. It means that the choice of a proper set of spectral lines is crucial for
element stratification study. Ideally, it should be a set of lines originated from the
levels in large range of the excitation energies and from different ions, 
thus probing a substantial part of stellar atmosphere. 
In chemically peculiar 
stars as cool as \hd\, we deal with severe blending by numerous lines of the rare-earth elements, 
therefore the strongest lines
with the developed wings (like, for example, \ion{Fe}{i}~$\lambda$\,4045.8~\AA\, line) are not possible to use. 
Also, \ion{Fe}{ii} lines with the upper level
excitation potential at $10$~eV and higher are not visible any longer. The blue spectral region is more crowded 
with the REE lines than the red one and the preferential working spectral region is $5000$--$6800$\AA\AA. 
Table~\ref{Tstrat-list} gives a list of Si, Ca, Cr, Fe, Sr, Ba lines used
in LTE stratification analysis routines together with the atomic parameters and the references. 
For NLTE stratification analysis we used the same lines as given in the papers of \citet{nd-NLTE} and \citet{pr-NLTE}.     

\begin{table*}[!th]
\caption{A list of spectral lines used for the stratification calculations. The columns give the ion identification,
central wavelength, the excitation potential (in eV) of the lower level, oscillator strength (\loggf), 
the Stark damping constant (``appr'' marks lines for which the classical approximation was used), and the reference 
for oscillator strength.}
\label{Tstrat-list}
\begin{footnotesize}
\begin{center}
\begin{tabular}{lcrrrl|lcrrrl}
\noalign{\smallskip}
\hline
\hline
Ion & Wavelength &\ei\,(eV)  &\loggf&$\log\,\gamma_{\rm St}$ & Ref.&Ion &Wavelength &\ei\,(eV)  &\loggf&$\log\,\gamma_{\rm St}$& Ref.\\
\hline
\ion{Si}{ii}&  5056.320& 10.074 & $-$0.510& $-$4.78 &BBC  &\ion{Cr}{ii}&  6147.154& 4.756 & $-$2.89 & $-$6.66 &RU  \\     
\ion{Si}{i} &  5517.530&  5.082 & $-$2.490& appr    &astr &            & 	  & 	  &         &	      &    \\   
\ion{Si}{i} &  5780.380&  4.920 & $-$2.350& $-$4.18 &G    &\ion{Fe}{ii}&  5132.669& 2.807 &$-$4.09  & $-$6.600&RU  \\   
\ion{Si}{i} &  5948.540&  4.930 & $-$1.230& $-$4.27 &G    &\ion{Fe}{ii}&  5169.033& 2.891 &$-$1.25  & $-$6.590&RU  \\   
\ion{Si}{ii}&  5978.930& 10.074 & $-$0.030& $-$5.01 &BBC  &\ion{Fe}{ii}&  5197.480& 5.960 &$-$3.13  & $-$6.700&RU  \\   
\ion{Si}{i }&  6142.480&  5.619 & $-$1.420& appr    &astr &\ion{Fe}{ii}&  5197.577& 3.230 &$-$2.10  & $-$6.600&RU  \\   
\ion{Si}{i }&  6155.130&  5.619 & $-$0.800& appr    &astr &\ion{Fe}{i }&  5198.711& 2.223 &$-$2.135 & $-$6.190&MFW \\     
\ion{Si}{ii}&  6347.110&  8.121 &    0.230& $-$5.31 &BBC  &\ion{Fe}{i} &  5217.389& 3.211 &$-$1.070 & $-$5.450&BKK \\   
\ion{Si}{ii}&  6371.350&  8.121 & $-$0.080& $-$5.32 &BBC  &\ion{Fe}{i} &  5253.462& 3.283 &$-$1.44  & $-$5.460&K07 \\      
            &          &        &         &         &     &\ion{Fe}{i} &  5383.369& 4.312 &   0.645 & $-$5.180&BWL \\     
\ion{Ca}{i} &  4226.728&  0.000 &    0.244& $-$6.03 & SG  &\ion{Fe}{i} &  5397.127& 0.915 &$-$1.993 & $-$6.300&MFW \\   
\ion{Ca}{ii}&  5021.138&  7.515 & $-$1.207& $-$4.61 & TB  &\ion{Fe}{i} &  5397.190& 4.446 &$-$1.16  & $-$5.260&K07 \\  
\ion{Ca}{ii}&  5285.138&  7.515 & $-$1.207& $-$4.61 & TB  &\ion{Fe}{i} &  5410.910& 4.470 &   0.398 & $-$5.060&BWL \\     
\ion{Ca}{i} &  5857.451&  2.933 &    0.240& $-$5.42 & S   &\ion{Fe}{i} &  5424.068& 4.320 &   0.520 & $-$4.790&MFW \\    
\ion{Ca}{i} &  5867.562&  2.933 & $-$1.570& $-$4.70 & S   &\ion{Fe}{ii}&  5425.257& 3.199 &$-$3.16  & $-$6.600&RU  \\    
\ion{Ca}{i} &  6162.173&  1.899 & $-$0.090& $-$5.32 & SO  &\ion{Fe}{i} &  5434.523& 1.011 &$-$2.122 & $-$6.303&BPS1\\   
\ion{Ca}{i} &  6163.755&  2.521 & $-$1.286& $-$5.00 & SR  &\ion{Fe}{i} &  5560.211& 4.434 &$-$1.05  & $-$4.323&K07 \\   
\ion{Ca}{i} &  6169.042&  2.253 & $-$0.797& $-$5.00 & SR  &\ion{Fe}{i} &  5576.089& 3.430 &$-$1.000 & $-$5.490&MFW \\   
\ion{Ca}{i} &  6455.598&  2.523 & $-$1.340& $-$6.07 & S   &\ion{Fe}{i} &  5775.081& 4.220 &$-$1.15  & $-$5.560&K07 \\   
\ion{Ca}{ii}&  6456.875&  8.438 &    0.410& $-$3.70 & TB  &\ion{Fe}{i} &  6137.691& 2.588 &$-$1.403 & $-$6.112&BPS2\\   
\ion{Ca}{i} &  6471.662&  2.526 & $-$0.686& $-$6.07 & SR  &\ion{Fe}{i} &  6336.824& 3.686 &$-$0.856 & $-$5.467&BK  \\  
            &          &        &         &         &     &\ion{Fe}{ii}&  6432.680& 2.891 &$-$3.52  & $-$6.690&RU  \\               
\ion{Cr}{i }&  4274.800&  0.000 & $-$0.231& $-$6.22 &MFW  &		          &	  &         &         &    \\ 
\ion{Cr}{ii}&  4592.050&  4.074 & $-$1.419& $-$6.65 &NL   &\ion{Sr}{ii}&  4161.792& 2.940 &$-$0.502 &  appr   &W   \\    
\ion{Cr}{ii}&  4634.070&  4.072 & $-$0.980& $-$6.65 &NL   &\ion{Sr}{ii}&  4215.519& 0.000 &$-$0.145 & $-$5.50*&W   \\    
\ion{Cr}{ii}&  5046.429&  8.227 & $-$1.75 & $-$5.91 &RU   &\ion{Sr}{i} &  4811.877& 1.847 &   0.190 &    appr &GC  \\     
\ion{Cr}{ii}&  5267.030&  4.042 & $-$3.06 & $-$6.72 &RU   &\ion{Sr}{i} &  5504.177& 2.259 &   0.090 &    appr &GC  \\     
\ion{Cr}{i} &  5296.691&  0.983 & $-$1.400& $-$6.12 &MFW  &\ion{Sr}{i} &  6408.459& 2.271 &   0.510 &    appr &GC  \\                                                            			  
\ion{Cr}{i} &  5297.377&  2.900 &    0.167& $-$4.31 &MFW  &            &          &       &         &         &    \\   
\ion{Cr}{i} &  5348.315&  1.004 & $-$1.290& $-$6.11 &MFW  &\ion{Ba}{ii}&  4166.000& 2.722 &$-$0.42  &    appr &NBS \\ 
\ion{Cr}{ii}&  5510.700&  3.827 & $-$2.610& $-$6.65 &RU   &\ion{Ba}{ii}&  4524.925& 2.512 &$-$0.36  &    appr &NBS \\                                                            
\ion{Cr}{ii}&  5569.110& 10.872 &    0.860& $-$5.36 &RU   &\ion{Ba}{ii}&  4554.029& 0.000 &   0.17  &    appr &NBS \\   
\ion{Cr}{ii}&  6053.466&  4.745 & $-$2.230& $-$6.63 &RU   &\ion{Ba}{ii}&  5853.668& 0.604 &$-$1.00  &    appr &NBS \\   
\ion{Cr}{ii}&  6070.100&  4.750 & $-$2.990& $-$6.63 &RU   &\ion{Ba}{ii}&  6141.713& 0.704 &$-$0.076 &    appr &NBS \\   
\ion{Cr}{ii}&  6138.721&  6.484 & $-$2.160& $-$6.73 &RU   &\ion{Ba}{ii}&  6496.897& 0.604 &$-$0.377 &    appr &NBS \\ 
\hline                                
\end{tabular}
\end{center}
G -- \citet{G73}; BBC -- \citet{BBC95}; 
SG -- \citet{SG}; SMP -- \citet{TB}); S -- \citet{S}); SO -- \citet{SN}; SR -- \citet{SR};
RU -- \citet{RU}; MWF -- \citet{MFW}; NL -- \citet{Cr2-Lund};
BWL -- \citet{BWL};
BPS1 -- \citet{BPS1}; 
BPS2 -- \citet{BPS2}; BK -- \citet{BK}; BKK -- \citet{BKK}; K07 -- (Kurucz database\footnote{\tt http://cfaku5.cfa.harvard.edu/ATOMS})
W -- \citet{W68}; GC -- \citet{GC88}; * -- \citet{Stark-Sr};
NBS -- \citet{NBS}.
\end{footnotesize}
\end{table*}

\begin{figure}
\includegraphics[width=\hsize]{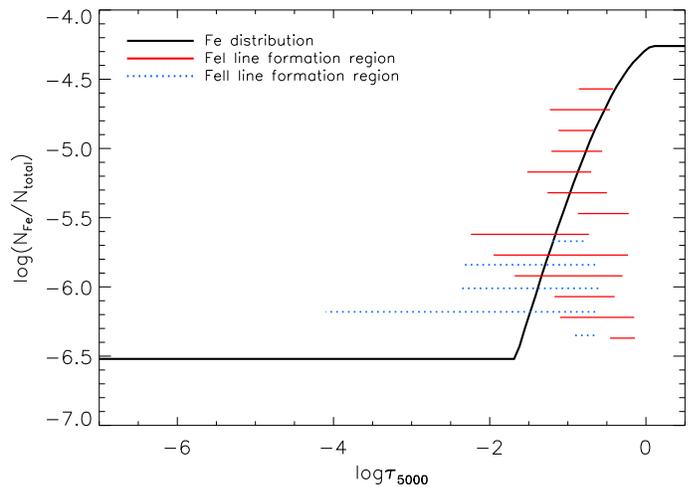}
\caption{Fe abundance distribution derived with the homogeneous model atmosphere and the range 
of the optical depth formation of \ion{Fe}{i} and \ion{Fe}{ii} lines from Table\,\ref{Tstrat-list} 
calculated with homogeneous Fe abundance $\log(Fe/N_{\rm total})=-5.10$.}
\label{Fdepth}
\end{figure}

\begin{figure*}
\includegraphics[width=\hsize]{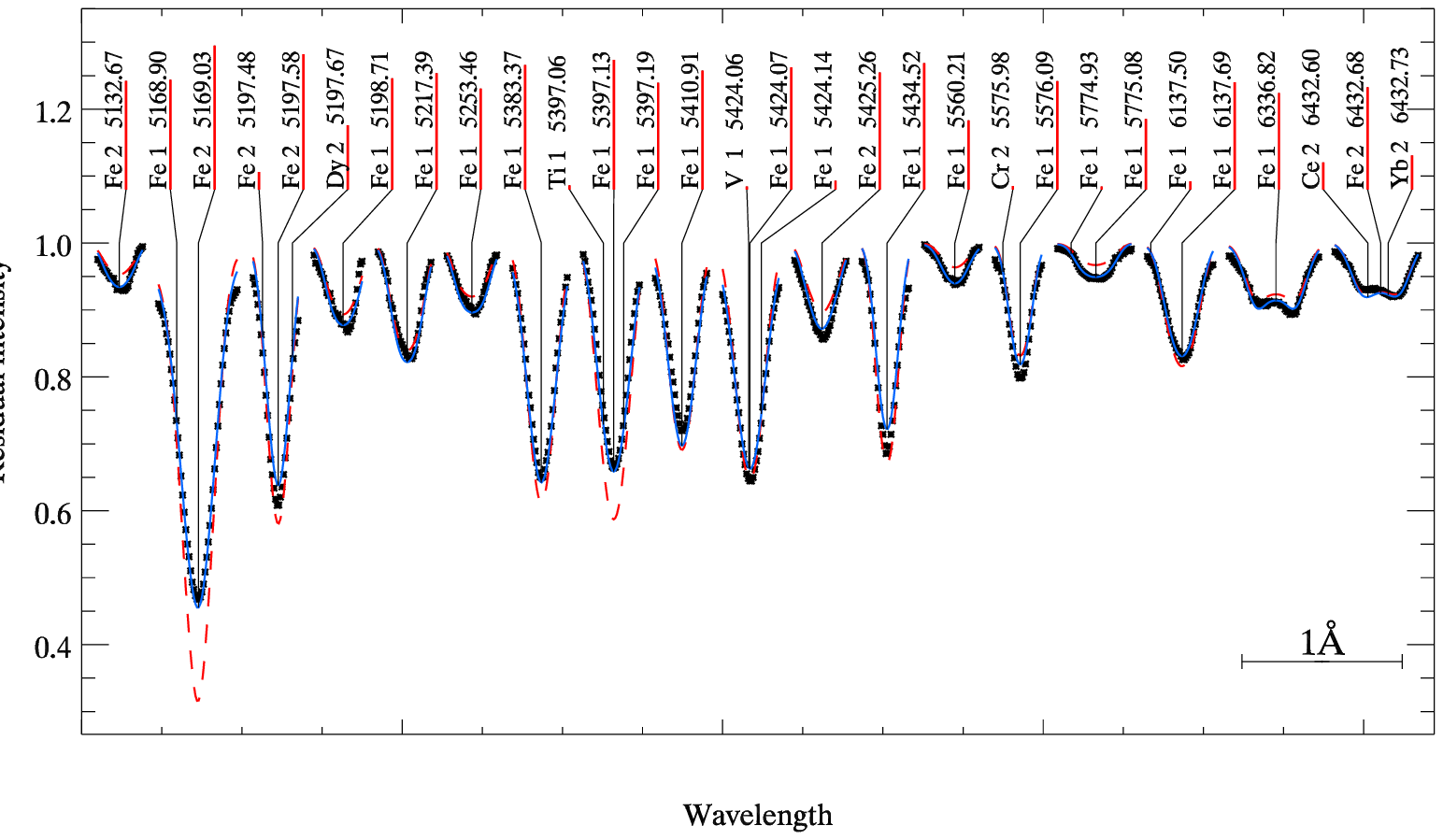}
\caption{A comparison between the observed Fe line profiles and calculations with
the stratified abundance distribution (full line) and with the homogeneous ($\log(Fe/N_{\rm total})=-5.10$) 
abundances (dashed line).}
\label{Fprofiles}
\end{figure*}

An ability of the chosen set of spectral lines to probe different atmospheric layers is illustrated in Fig.\,\ref{Fdepth}.
The iron stratification derived with the initial model (see Section\,\ref{sres}) is shown together with the ranges
of Fe lines depth formation calculated on $\taunu$ scale. The later were calculated with the contribution 
function to the emergent line radiation following \citet{Achmad91}. Fig.\,\ref{Fprofiles} illustrates a fit of the calculated
to the observed \ion{Fe} line profiles for homogeneous and stratified atmospheres. For Si, Ca, 
Cr, Sr, and Ba the corresponding plots are shown in Figs.\,\ref{Si},\ref{Ca},\ref{Cr},\ref{Sr-Ba} (Online material).

\subsection{Self-consistent models with empirical stratification}
Using the methods of stratification analysis described above one could restore the element distribution
profile of any chemical elements for which accurate atomic line data exist.
However, we should remember that the empirical analysis of chemical elements
stratification is based on model atmosphere technique, 
and thus the temperature-pressure structure of model atmosphere itself depends upon
stratification which has to be found. Therefore, the calculation of model atmosphere
and stratification (abundances) analysis are linked together
and thus an iterative procedure should be used in this case. 
In general, this procedure could be divided into several steps:

\begin{enumerate}
\item
calculation of so-called first approximation, simplified model, 
that could be any standard model atmosphere with the atmospheric parameters
close to those of the star to be analysed;
\item
determination of stratification of chemical elements based on spectroscopic analysis of individual spectral lines;
\item
calculation of improved model atmosphere taking into account stratification found in previous step;
\item
comparison of modeled energy distribution (or/and photometric colors) and hydrogen line profiles 
with observed ones and re-adjustments of model input parameters ($\teff$, $\logg$) 
if needed to fit observations. At this step several calculations of model atmosphere with fixed stratification 
but different model parameters may be required;
\item
repeating the overall process starting from step~2 until stratification profiles of chemical elements 
and model parameters are not converged.
\end{enumerate}

At the end, the listed above procedure ensures the consistency between model atmosphere structure 
and abundances used for calculation of synthetic line profiles and interpretation of observed data.

\section{Results}\label{sres}
\subsection{Construction of model atmosphere}
In the present analysis we started from chemically homogeneous  model atmosphere with the parameters
$\teff=7250$\,K and $\logg=4.3$ and the abundances taken from \citet{ryabchik97} (Model 1).
These mean abundances were derived based on the observations close to magnetic maximum,
and they provide a satisfactory fit for HARPS observations. Table~\ref{Tabn} shows abundances of the elements which were not passed
through the subsequent stratification analysis.  
For the rest of chemical elements the solar composition \citep{asplund} was assumed.
  
\begin{figure}
\includegraphics[width=\hsize]{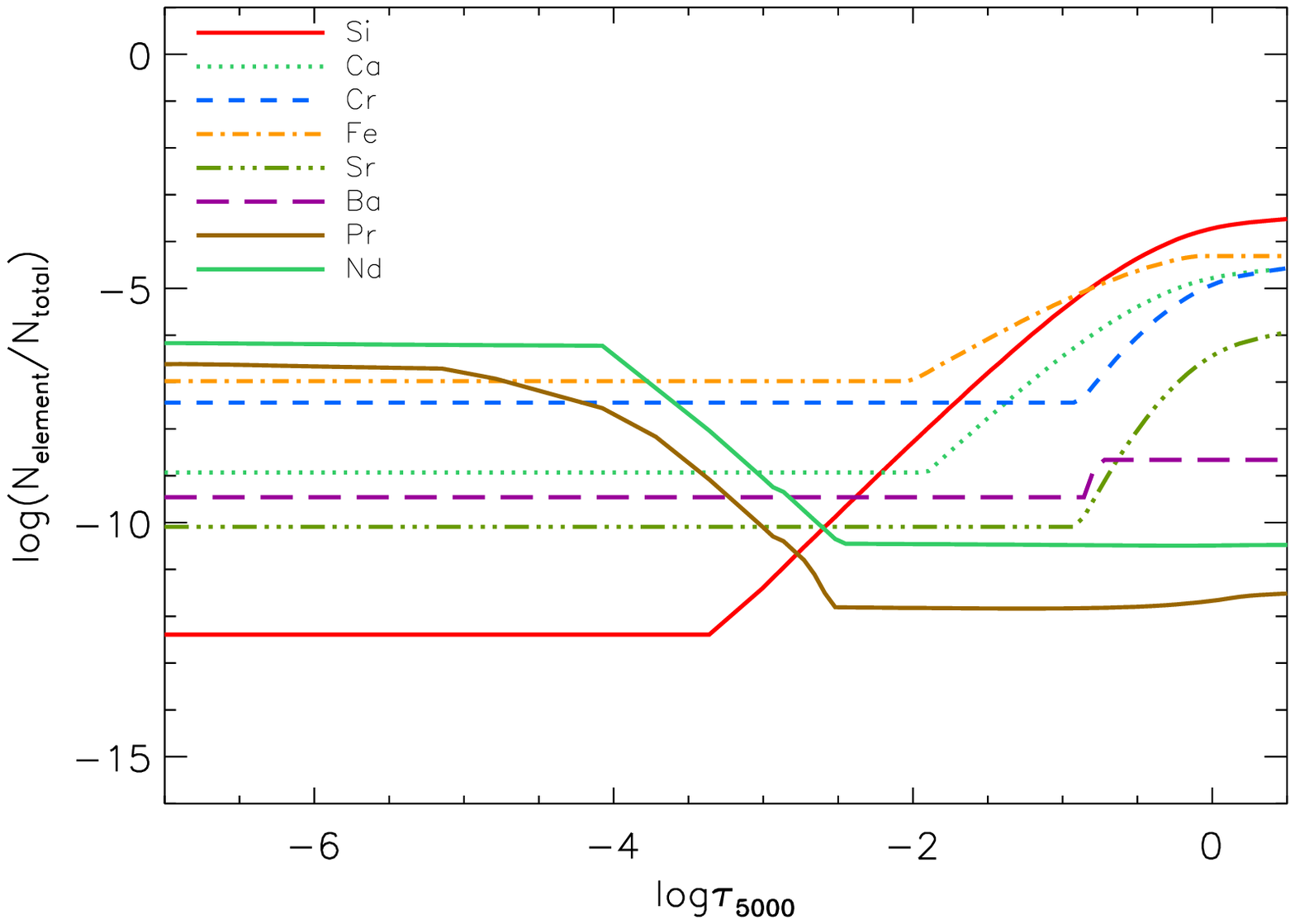}
\includegraphics[width=\hsize]{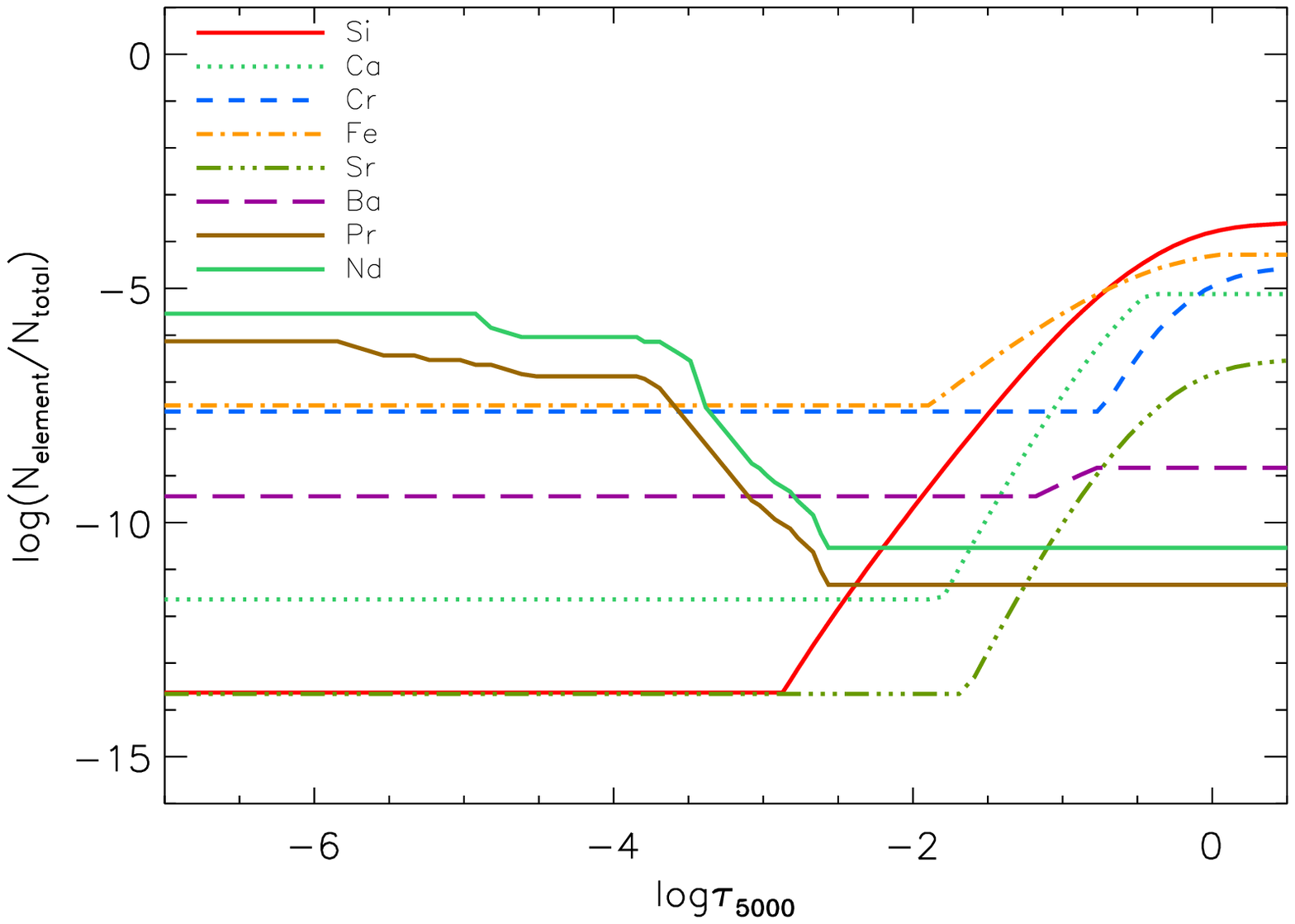}
\caption{Stratification of eight elements in atmosphere of \hd\, 
derived using homogeneous first-approximation model (Model~1, top panel) and
using final model with scaled REE opacity (Model~5, bottom panel).}
\label{Fstrat}
\end{figure}

Based on this model the stratification of 
Fe, Ba, Ca, Sr, Cr, Si, Pr, and Nd was derived and is displayed on the top panel of Fig.~\ref{Fstrat}.
It is seen that the stratification of majority of elements shows element underabundance 
in surface layers and
overabundance in layers around star's photosphere ($\log\tau_{\rm 5000}\approx[-1,0]$). 
Contrary, the distribution of Pr and Nd show inverse picture with the strong overabundance (up to $5$\,dex) in surface layers.
This kind of elements distribution, i.e. the difference between distribution of REE's comparing to other elements, 
seems to be a common characteristic of the atmospheres  of Ap stars \citep{str3}.

The stratification shown in Fig.~\ref{Fstrat} (top panel) was then applied to the calculation of new model atmosphere.
Again, the individual values of abundances of some other elements used for calculation were taken from Table~\ref{Tabn}. 

\begin{table}
\caption{Abundances of individual elements.}
\begin{center}
\begin{tabular}{ccc}
\hline\hline
Element & $\log(N_{\rm el}/N_{\rm total})$ & $\log(N_{\rm el}/N_{\rm total})_{\bigodot}$\\
\hline
\ion{C}    & $-4.00$  & $ -3.65$\\
\ion{N}    & $-4.50$  & $ -4.26$\\
\ion{O}    & $-4.00$  & $ -3.38$\\
\ion{Mg}   & $-5.60$  & $ -4.51$\\
\ion{Al}   & $-5.57$  & $ -5.67$\\
\ion{Ti}   & $-7.28$  & $ -7.14$\\
\ion{Mn}   & $-7.01$  & $ -6.65$\\
\ion{Co}   & $-5.57$  & $ -7.12$\\
\ion{Ni}   & $-6.54$  & $ -5.81$\\
\ion{Y}    & $-7.80$  & $ -9.83$\\
\ion{La}   & $-9.00$  & $-10.91$\\
\ion{Ce}   & $-8.90$  & $-10.46$\\
\ion{Sm}   & $-9.00$  & $-11.03$\\
\ion{Eu}   & $-9.30$  & $-11.52$\\
\ion{Gd}   & $-8.70$  & $-10.92$\\
\ion{Dy}   & $-8.94$  & $-10.90$\\
\ion{Er}   & $-9.53$  & $-11.11$
\end{tabular}
\label{Tabn}
\end{center}
\smallskip

The abundances derived from observations are from \citet{ryabchik97}. 
The REE's abundances were derived using the lines of the first ions. The solar abundances are taken from \citet{asplund}.
\end{table}

\begin{figure}
\includegraphics[width=\hsize]{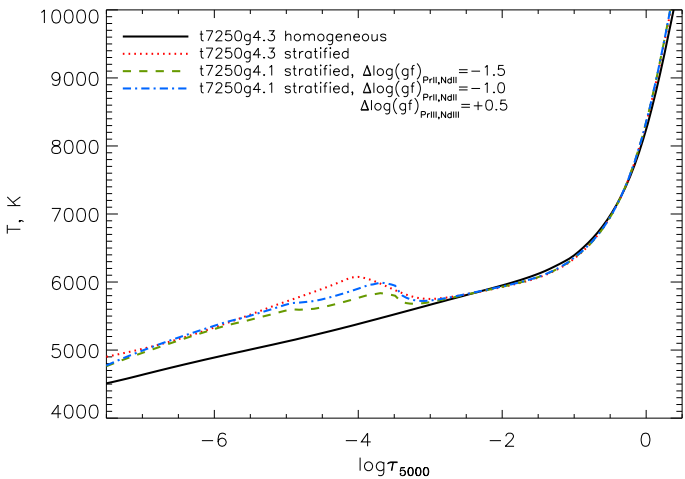}
\caption{Effect of Pr and Nd stratification on model temperature structure. 
Solid line~--~t7250g4.3 model calculated with
homogeneous abundances; dotted line~--~t7250g4.3 model calculated with
stratified abundances shown in Fig.~\ref{Fstrat}; 
dashed and dash-dotted lines~--~t7250g4.1 models calculated with stratified abundances and 
scaled \ion{Pr}{ii,iii} and \ion{Nd}{ii,iii} \loggf\ values.}
\label{Ftjump}
\end{figure}

The stratification of REE elements and its implementation in model atmosphere computation 
deserve more detailed description.
Numerical calculations showed that the REE elements stratification 
has a strong influence of model atmosphere
structure leading to the appearance of inverse temperature gradient in upper atmospheric region. 
This is illustrated in Fig.~\ref{Ftjump},
where the models with the same atmospheric parameters and input line list were computed 
with and without stratification of chemical elements
(solid and dotted lines respectively). 
It is seen that the heating of the stratified atmosphere starts at $\log\tau_{\rm 5000}=-3$ going outwards
to the stellar surface and reaches its maximal of about $800$\,K around $\log\tau_{\rm 5000}=-4$ 
compared to homogeneous abundance model.
This temperature increase is caused by the presence of REE stratification and, 
as it seen from  Fig.~\ref{Fstrat}, is directly related
to the position of abundance jumps of Pr and Nd. 
No other elements considered in our computations are responsible for such an effect.
Indeed, the strong overabundance of Pr and Nd by $4$--$5$\,dex relative to their solar values 
and their rich spectra presented in
current version of \vald\ database make the absorption coefficient to be very high at frequencies of REE transitions.
The latter are generally located at the spectral region where the star with $\teff=7250$\,K 
radiates most of its energy dominating
in the total radiative energy balance, 
and thus allowing for the heating of atmospheric layers with high REE opacity.

The appearance of inverse temperature gradient clearly shows that the REE elements 
are no longer trace elements in atmospheres of
Ap stars if their stratification is as strong (or comparably strong) as in the atmosphere of \hd. 
This immediately implies that not only synthetic
line profiles, but also model atmospheres with REE stratification must be computed 
based on NLTE line formation for REE's.
Unfortunately, this can not be done with the current version of \llm.
Nevertheless, to simulate NLTE effects of Pr and Nd in model atmosphere calculations 
we used the following approach. 
\citet{nd-NLTE} and \citet{pr-NLTE} found that, for the stratified distribution of Nd and Pr, the NLTE abundance corrections are 
positive for the lines of the first ions at a level of one order of magnitude and negative for the lines of the second ions at a level of 
$\simeq0.5$~dex.
Thus, to account for the REE stratification 
in LTE model atmosphere code, we scale the \loggf\, values of \ion{Pr}{ii} and \ion{Nd}{ii} lines by $-1.5$\,dex
while taking abundances derived using second 
ions for model atmosphere calculation.
This procedure was 
applied to all the  \ion{Pr}{ii} and \ion{Nd}{ii} lines that are presented in master line list
used for model atmosphere calculations. Then, using this new line list, we recomputed the model atmosphere 
and re-derived stratification of chemical elements again. 
At each iteration Pr and Nd stratifications were recalculated 
based on NLTE line formation for \ion{Pr}{ii/iii}\, and \ion{Nd}{ii/iii}.
The best fit to hydrogen \halpha\, line profile (wings)
and photometric color-indices required the decrease of effective gravity down to $\logg=4.1$ 
(mainly to fit $c_{\rm 1}$ index). No further noticeable changes
were detected in observed parameters and this gravity 
was considered as a final one. The model and stratification
calculations are converged after the second iteration (Model 4).

Except Pr and Nd, other REE elements can also be non-uniformly distributed in the atmosphere of \hd\, 
and can also exhibit deviations from LTE. Among them, only most abundant elements may play 
a noticeable role in modification of atmospheric structure. At present, we do know neither distributions 
of these elements, nor the magnitude of their NLTE effects.
In any case, preliminary abundance results for Ce and Sm show that REE anomaly -- a discrepancy between abundances 
inferred from the lines of singly- and doubly-ionized element -- reaches approximately the same magnitude 
as for Pr and Nd. Supposing that the Ce (which has the biggest number of available lines in \vald\, 
among other REE elements) has the same vertical distribution as Nd, we estimated its effect on atmospheric structure. 
We found that neglecting Ce stratification we underestimate a size of temperature jump by no more than $\sim$200~K.

Decreasing the \loggf\ values of \ion{Pr}{ii} and \ion{Nd}{ii}
lines could, in principle, simulate the NLTE line opacity of these elements, however the second ions
are also affected by NLTE effects that bring as much as $0.5$\,dex difference between LTE and
NLTE abundances derived using second ions spectra. This forced us to compute another model
where the \loggf\ values of both first and second ions are changed by $-1$\,dex and $0.5$\,dex
respectively (Model 5). This model provides slightly higher temperature increase in superficial layers 
compared with the model where \loggf\ values of only first ions were changed (see Fig.~\ref{Ftjump}).
It is seen that the simulation of NLTE REE opacity tends to decrease the amplitude of the 
temperature jump almost by a factor of two. 
However, independently on the way how to scale the REE opacity,
the overall picture remains the same illustrating the heating of surface layers in the region 
of strong Pr and Nd overabundance.

\begin{table*}[!t]
\caption{Observed and calculated photometric parameters of \hd.}
\begin{footnotesize}
\begin{center}
\begin{tabular}{c|c|ccccc}
\hline\hline
                &                                                    & t7250g4.3   & t7250g4.3     & t7250g4.1	   & t7250g4.1				   & t7250g4.1\\
Color           &                                                    & homogeneous & stratified    & stratified    & stratified			           & stratified\\
index           &      \textbf{Observations}                         &		  & REE  unscaled & REE  unscaled & $\Delta$\loggf$_{\rm PrII,NdII}=-1.5$ & $\Delta$\loggf$_{\rm PrII,NdII}=-1$\\
                &       	                                     &		  &		  &		  &					  & $\Delta$\loggf$_{\rm PrIII,NdIII}=+0.5$\\
                &       	                                     & (Model 1)   & (Model 2)	  & (Model 3)	  & (Model 4)				  & (Model 5)\\
\hline
$b$-$y$         &$\mathbf{0.186}~($0.003$)$;~~~$\mathbf{0.191}^\star$&$0.195$      &$0.188$        &$0.183$        &$0.180$                                &$0.180$\\
$m_{\rm 1}$     &$\mathbf{0.202}~($0.006$)$;~~~$\mathbf{0.211}^\star$&$0.197$	   &$0.214$	   &$0.213$	   &$0.205$				   &$0.207$\\
$c_{\rm 1}$     &$\mathbf{0.653}~($0.011$)$;~~~$\mathbf{0.626}^\star$&$0.610$	   &$0.575$	   &$0.638$	   &$0.663$				   &$0.656$\\
$H\beta$        &$\mathbf{2.745}~($0.003$)$;~~~$\mathbf{2.760}^\star$&$2.802$	   &$2.804$	   &$2.808$	   &$2.810$				   &$2.809$\\
$a$             &$\mathbf{0.609}~($0.0005$)$      		     &$0.621$	   &$0.621$	   &$0.621$	   &$0.617$				   &$0.619$\\
$B$-$V$         &$\mathbf{0.320}$       			     &$0.308$	   &$0.308$	   &$0.297$	   &$0.288$				   &$0.290$\\
$U$-$B$         &$\mathbf{1.381}$       			     &$1.333$	   &$1.311$	   &$1.370$	   &$1.379$				   &$1.376$\\
$V$-$B$         &$\mathbf{0.572}$       			     &$0.596$	   &$0.595$	   &$0.610$	   &$0.621$				   &$0.619$\\
$B_{\rm1}$-$B$  &$\mathbf{0.978}$       			     &$0.984$	   &$0.990$	   &$0.985$	   &$0.982$				   &$0.983$\\
$B_{\rm2}$-$B$  &$\mathbf{1.393}$       			     &$1.413$	   &$1.407$	   &$1.413$	   &$1.417$				   &$1.416$\\
$V_{\rm1}$-$B$  &$\mathbf{1.286}$       			     &$1.319$	   &$1.319$	   &$1.333$	   &$1.343$				   &$1.341$\\
$G$-$B$         &$\mathbf{1.688}$       			     &$1.727$	   &$1.726$	   &$1.741$	   &$1.753$				   &$1.751$\\
\hline
\end{tabular}
\end{center}
\label{Tcolors}
\smallskip
$^\star$~--~\citet{martinez}\\
The values in brackets give the error bars of observations.
\end{footnotesize}
\end{table*}

We also tested the influence of \ion{Pr}{ii} and \ion{Nd}{ii} bound-free opacity on model
structure and energy distribution. To simplify the computations we assumed three characteristic
energy levels with excitation energies of $2$~eV, $3$~eV, and $4$~eV for both ions. These levels further
consist of $50$ atomic states each with the statistical weight $g=7$. Such an approximation is based 
on the fact that both ions have essentially the same structure of states with the very close ionization potentials. 
The NLTE level populations were taken directly from \detail\ code and then
incorporated in \llm. For all the levels, the photoionization cross-sections were calculated using the hydrogenic approximation.
It appeared that the effect of the \ion{Pr}{ii} and \ion{Nd}{ii} bound-free opacity becomes visible
only if the abundances of these elements are increased by $+6$~dex for the stratified model 
where REE's are already in strong overabundance in surface layers. Even so, the cumulative effect
on on the temperature distribution was never higher than tens of K.
We, thus, concluded that REE bound-free opacity can be ignored in the model atmosphere computations.

Bottom panel of Fig.~\ref{Fstrat} shows the stratification derived with Model 5. 
One can note that the stratification of REE's shows 
more steep abundance jump in this case which is probably a result of more refined set of optical depths
around REE jump implemented in the last iteration compared to the initial model and the first iteration.
This is also the reason why the inverse temperature gradient shown in Fig.~\ref{Ftjump} is more steep for the
model calculated with Model 5 stratification than for the model with stratification derived in Model 4. 
Few remarks concerning Sr and Ba stratification calculations should be made. Although we get formally Ba stratification, it is not
significant because the difference between $\chi^2$ for line profiles fits with uniform and stratified models 
does not exceed 15-16\%, while $\chi^2$ differs by 5-10 times in the case of real stratification. Therefore uniform Ba distribution in
the atmosphere of \hd\ or at least in the layers where the studied \ion{Ba}{ii} lines are formed is more probable. For Sr distribution
we used five lines with three weakest lines of \ion{Sr}{i}. Strontium distribution
is mainly based on strong \ion{Sr}{ii} lines, in particular, on resonance \ion{Sr}{ii}~$\lambda$~4215.5~\AA. 
The wings of \ion{Sr}{ii} lines that probe abundance in the deeper atmospheric layers, are blended, that obviously leads to 
larger formal computing errors of stratification
parameters determination and, hence, to worse $\chi^2$. To check a reliability of Sr stratification we recalculated it using Sr lines
observed at slightly different magnetic phase $0.944$ (see UVES time-series observations in \citet{puls}). Within the formal
errors of \ddafit\, procedure strontium abundance profiles coincide  below  $\log\tau_{\rm 5000}=-2$  where practically 
all considered Sr lines are formed. They differ above $\log\tau_{\rm 5000}=-2$, where the only core of \ion{Sr}{ii}~$\lambda$~4215.5~\AA\ 
line is formed.
Poor fit of this line in the transition region between the line core and the line wings indicates that a more complex shape of
abundance profile than a simple step-function is required to describe the observed line profile, similar to
the Ap star HD\,133792 \citep{vip}. Strontium stratification derived for \hd\ is not spurious, but
the shape of the Sr abundance profile may differ from that derived in the present study.    
The other elements that demonstrated
strong changes during iterative stratification analysis are Si and Ca. 
Both elements show the increase of the abundance jump
amplitude in the final stratification with strong underabundance in surface layers and with practically the same abundance near
photospheric layers. The jump itself becomes steeper. 
It may be caused by the change in temperature and electron density structure of the atmosphere above $\log\tau_{\rm 5000}=-3$, where
the central parts of the strongest \ion{Ca}{i}~$\lambda$~4227 and \ion{Si}{ii}~$\lambda\lambda$~6347,6371 are formed. However,
the formal errors of Si and Ca abundances in the upper atmospheric layers derived in \ddafit\, procedure are rather high, up to 1.5 dex,
hence 2.5 dex difference for the initial and final Ca distributions may be considered as within the $3\sigma$ limit. 

The changes in element distributions between the two cases shown
in Figs.~\ref{Fstrat} are purely due to the changes in model atmosphere structure. Thus
we can conclude that the recalculation of the model atmosphere structure
could be very important for the quantitative stratification analysis. 
For some elements, the stratification derived with models computed under the assumption of 
homogeneous elements distribution can be considered only as a first approximation.

Finally, in Table~\ref{Tcolors} we compare theoretical and observed photometric color-indices of \hd.
The calculations are presented for two models with $\teff=7250$\,K, $\logg=4.3$ (homogeneous and stratified abundances) and for 
three models with $\teff=7250$\,K, $\logg=4.1$ but with different
assumptions about Pr and Nd opacity. 
The theoretical colors were calculated with
modified computer codes taken from \citet{a9}.
One can see that the color-indices of Str\"omgren system as well as $U-B$ and $V-B$ indices of Geneva system
are better represented by stratified models .
In contrast, the last two Geneva indices given in Table~\ref{Tcolors} are better described
by homogeneous abundance model.
However, these differences are not very big and we can conclude that,
generally, the two final models calculated with the complex stratification shown in Fig.~\ref{Fstrat} (bottom panel)
and scaled REE opacity demonstrates a good agreement with the observations.
We have to note that the Str\"omgren photometry taken from \citet{martinez}, which is not available
in electronic format, introduces additional uncertainties. In particular, index $b-y$ of \citet{martinez}
is better represented by homogeneous model. Furthermore, the difference of $0.027$\,mag between the
two observations for $c_{\rm 1}$ index (see Table~\ref{Tcolors}) does not allow to distinguish definitely between 
homogeneous and stratified models. 
On the other hand, the observation of $c_{\rm 1}$ taken from \citet{hauck} 
agrees nicely with predictions 
by model with scaled \ion{Pr}{ii,iii} and \ion{Nd}{ii,iii} opacity. 
The difference of $0.007$\,mag between models computed under two assumptions about 
the scaling of REE opacity (last two models in Table~\ref{Tcolors}) 
is within the error bars of observations, but deviates significantly from $c_{\rm 1}$ calculated
without REE scaling. Our calculations demonstrate that taken stratification effects into account we obtain lower surface
gravity than follows from the calibrations based on homogeneous models.  
Strong impact of simulated NLTE effects of Pr and Nd on $c_{\rm 1}$ value, temperature distribution,
and colors demonstrates a necessity of more precise model atmosphere computations with 
accurate treatment of Pr, Nd, and possibly other REE lines formation.
Our choice of $\logg$ for the final model could also change as more
advanced NLTE model atmospheres are employed. Moreover, the strong variability in 
$c_{\rm 1}$ index with the  peak-to-peak amplitude of $0.086$\,mag~\citep{wollf} 
is likely to be responsible for the difference between the value of \citet{martinez} 
and the average value given by \citet{hauck}.
Thus, at present it is difficult
to say something more definite about the $\logg$ value of the star and we restrict
ourselves with $\logg=4.1$.

\begin{figure}
\includegraphics[width=\hsize]{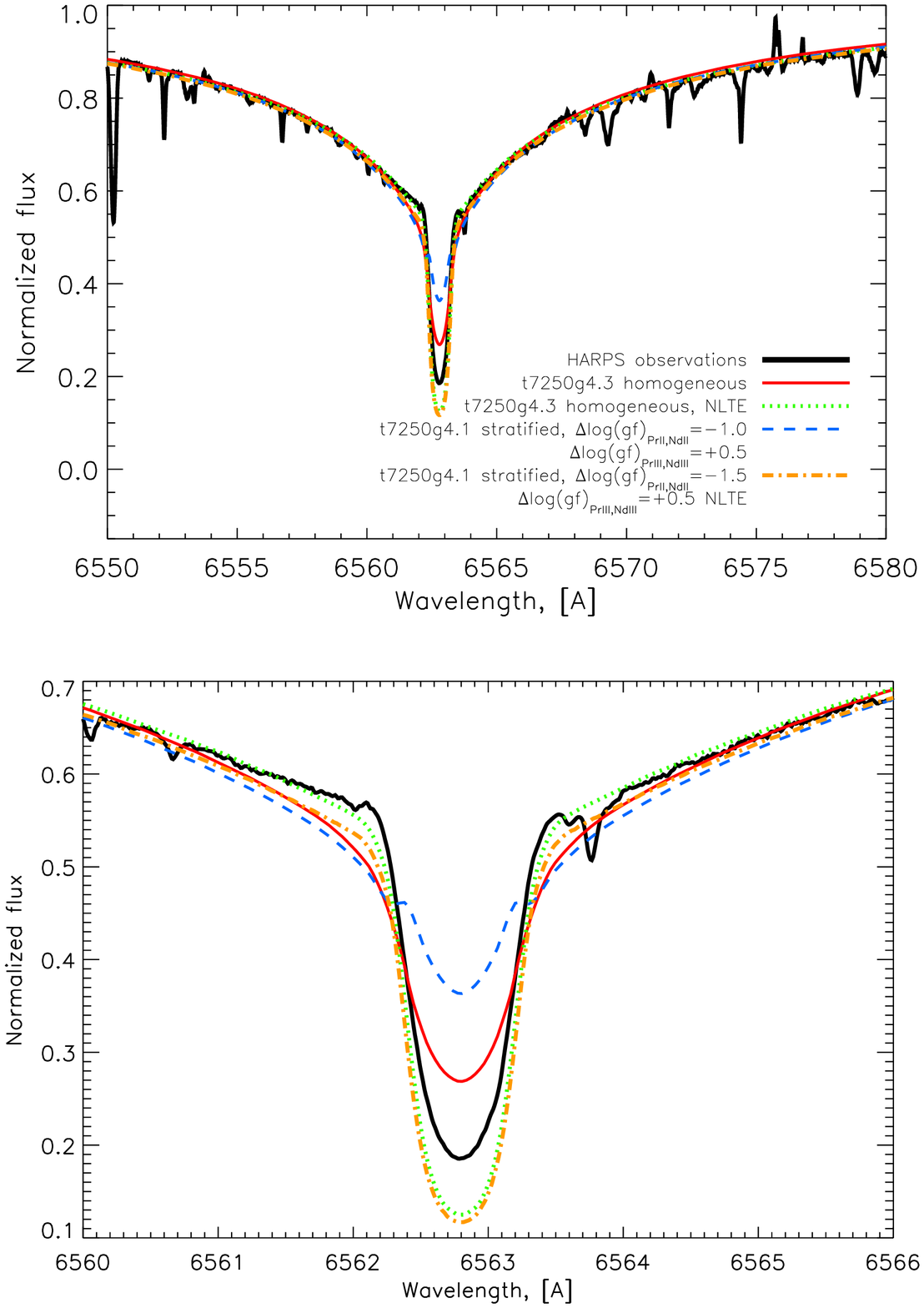}
\caption{Observed and calculated \halpha\, line profiles. Thick full line~--~HARPS observations; dotted line~--~model
with $\teff=7250$\,K, $\logg=4.3$ and with homogeneous elements distribution; full line~--~model
with $\teff=7250$\,K, $\logg=4.3$ and with stratification shown in~\ref{Fstrat} (top panel); dashed line~--~model
with $\teff=7250$\,K, $\logg=4.1$ and with final iteration shown in~\ref{Fstrat} (bottom panel). 
Bottom panel shows the zoomed part of the
figure on top panel around the core of the line.}
\label{Falpha}
\end{figure}

As it was stated above, for the opacity calculation in model atmospheres
we used REE data extracted from \vald\, database. However, for the NLTE calculations with \detail\, code
it was necessary to have more extensive line lists for accurate calculations of radiative rates for every single
atomic state. These detailed line lists for \ion{Pr}{ii} and \ion{Nd}{ii/iii} were produced in 
Institute of Spectroscopy (ISAN, Russia), and described in \citet{nd-NLTE,pr-NLTE}. 
Note that the ISAN data already contains Pr and Nd parameters from \vald\ thus providing only additional
number of predicted lines. For example, while \vald\ provides information about $508$ \ion{Pr}{ii},
$1279$ \ion{Nd}{ii}, and $55$ \ion{Nd}{iii} lines, ISAN theoretical computations extended them up to
$103428$, $1172579$, and  $6858$ lines respectively. However, this great increase 
in line numbers showed only marginal changes in model temperature
distribution. This allowed to conclude that the radiative equilibrium balance
in atmosphere of \hd\ is controlled by strongest REE lines that are presented in \vald.
The only noticeable changes were detected for the $c_{\rm 1}$ (increased by $0.012$~mag)
and for $m_{\rm 1}$ index (decreased by $0.008$~mag).

\begin{figure*}
\begin{center}
\includegraphics*[angle=90,width=\hsize]{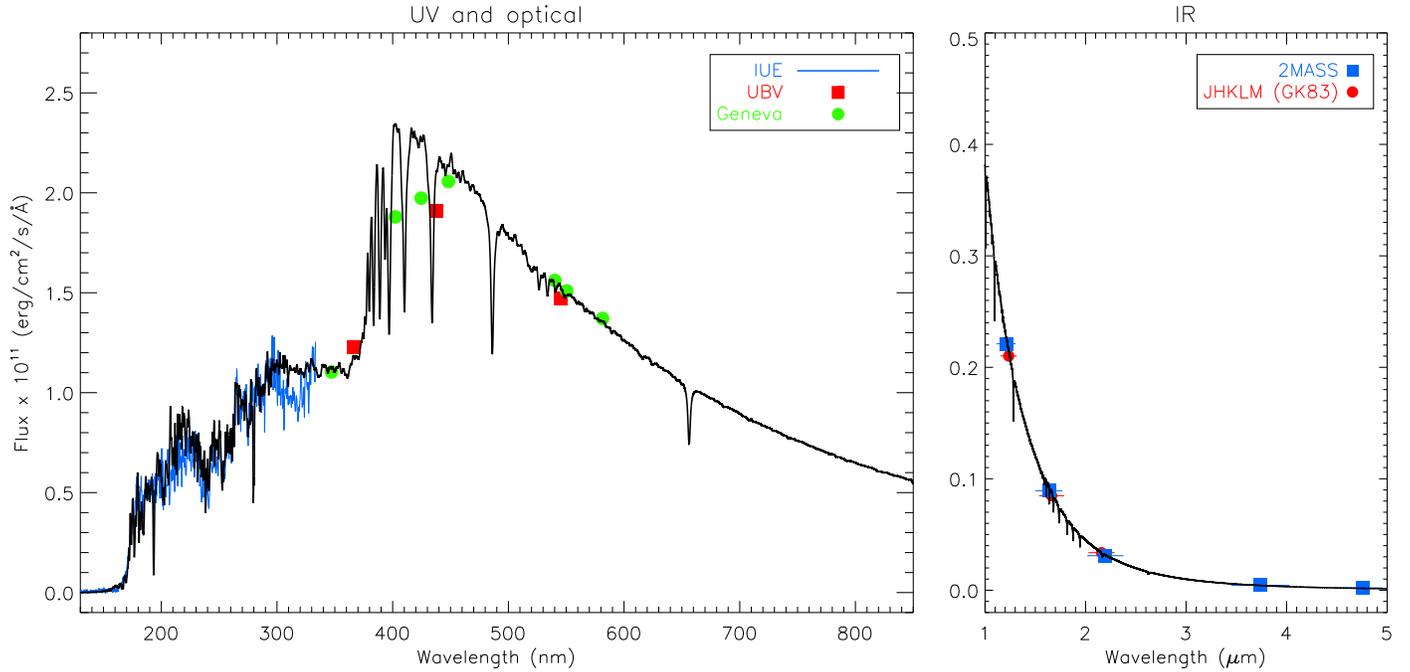}
\end{center}
\caption{Calculated and observed energy distributions of \hd. Solid line~--~t7250g4.1 model calculated with 
stratified abundances shown in Fig.~\ref{Fstrat} (bottom panel)
and changed \ion{Pr}{ii,iii} and \ion{Nd}{ii,iii} \loggf\ values by $-1$\,dex and $+0.5$\,dex respectively.}
\label{Fsed}
\end{figure*}

\begin{figure*}
\includegraphics[width=\hsize]{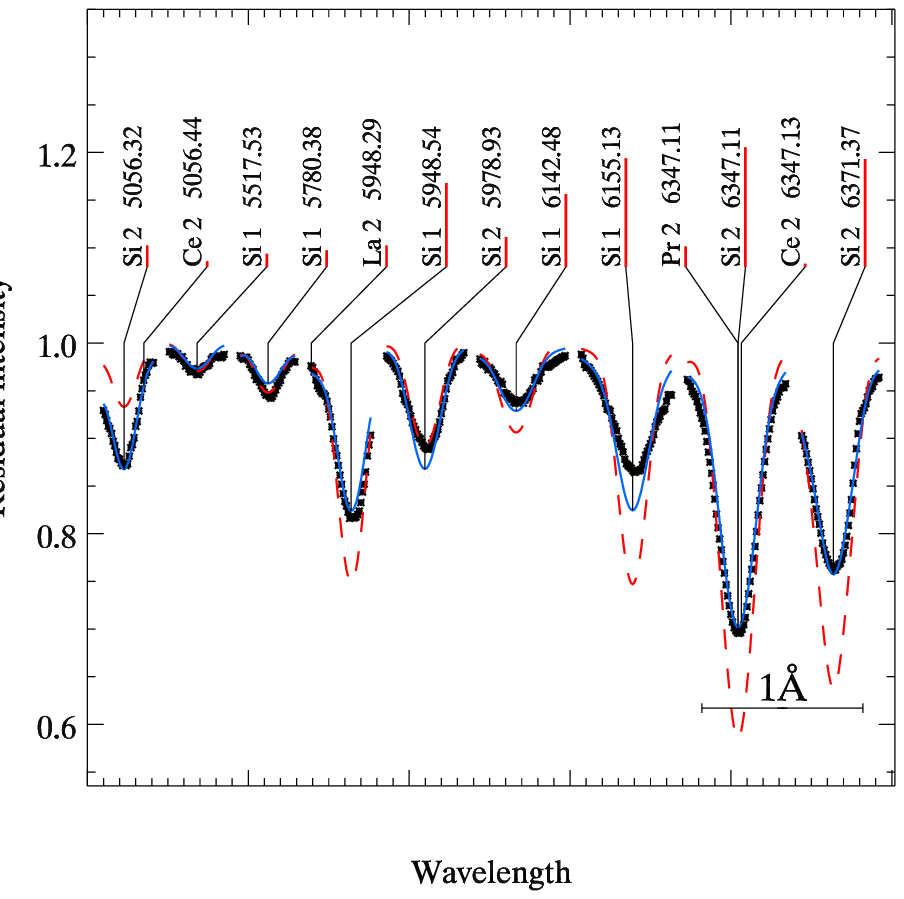}
\caption{A comparison between the observed \ion{Si} line profiles and calculations with
the stratified abundance distribution (full line) and with the homogeneous ($\log(Si/N_{\rm total})=-4.50$) 
abundances (dashed line).}
\label{Si}
\end{figure*}

\begin{figure*}
\includegraphics[width=\hsize]{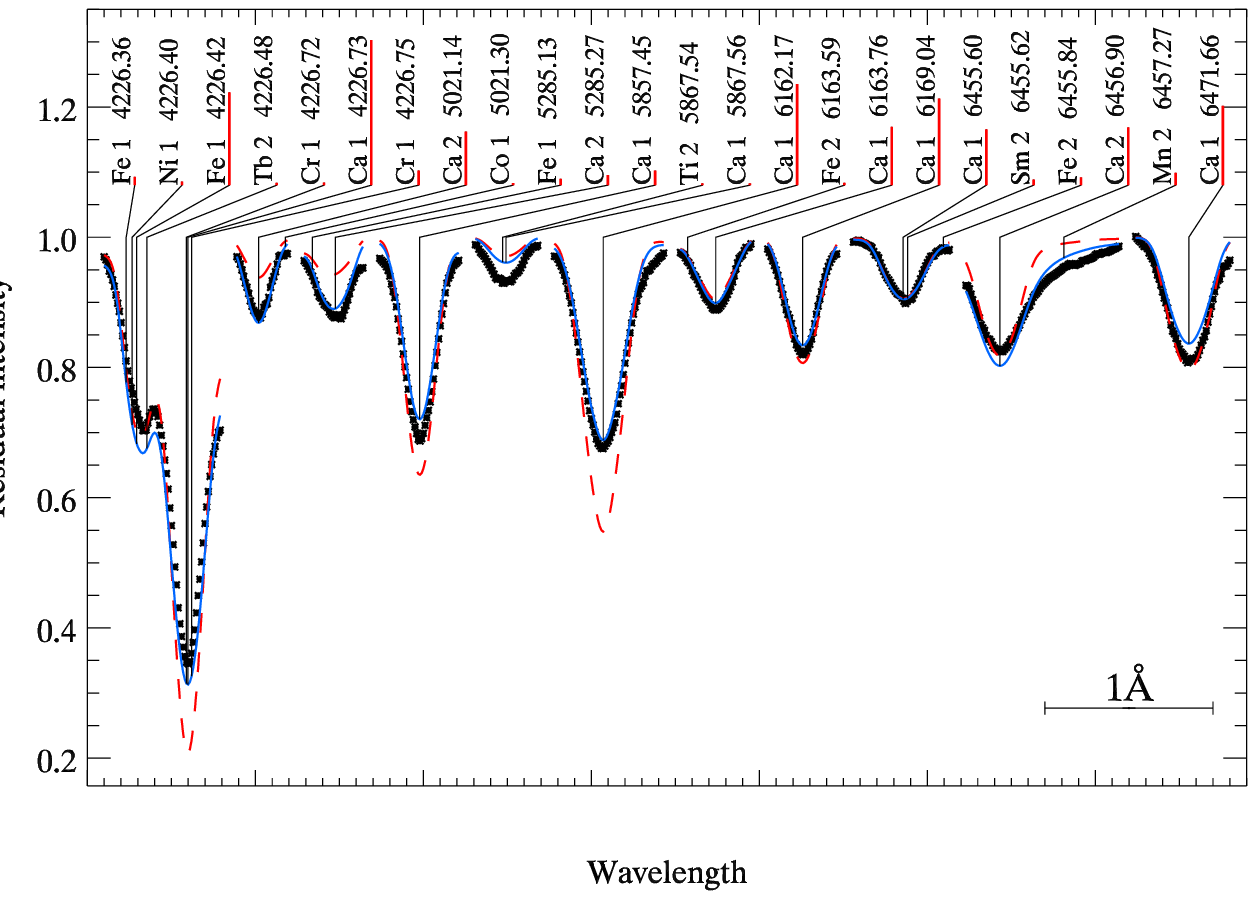}
\caption{A comparison between the observed \ion{Ca} line profiles and calculations with
the stratified abundance distribution (full line) and with the homogeneous ($\log(Ca/N_{\rm total})=-5.80$) 
abundances (dashed line).}
\label{Ca}
\end{figure*}

\begin{figure*}
\includegraphics[width=\hsize]{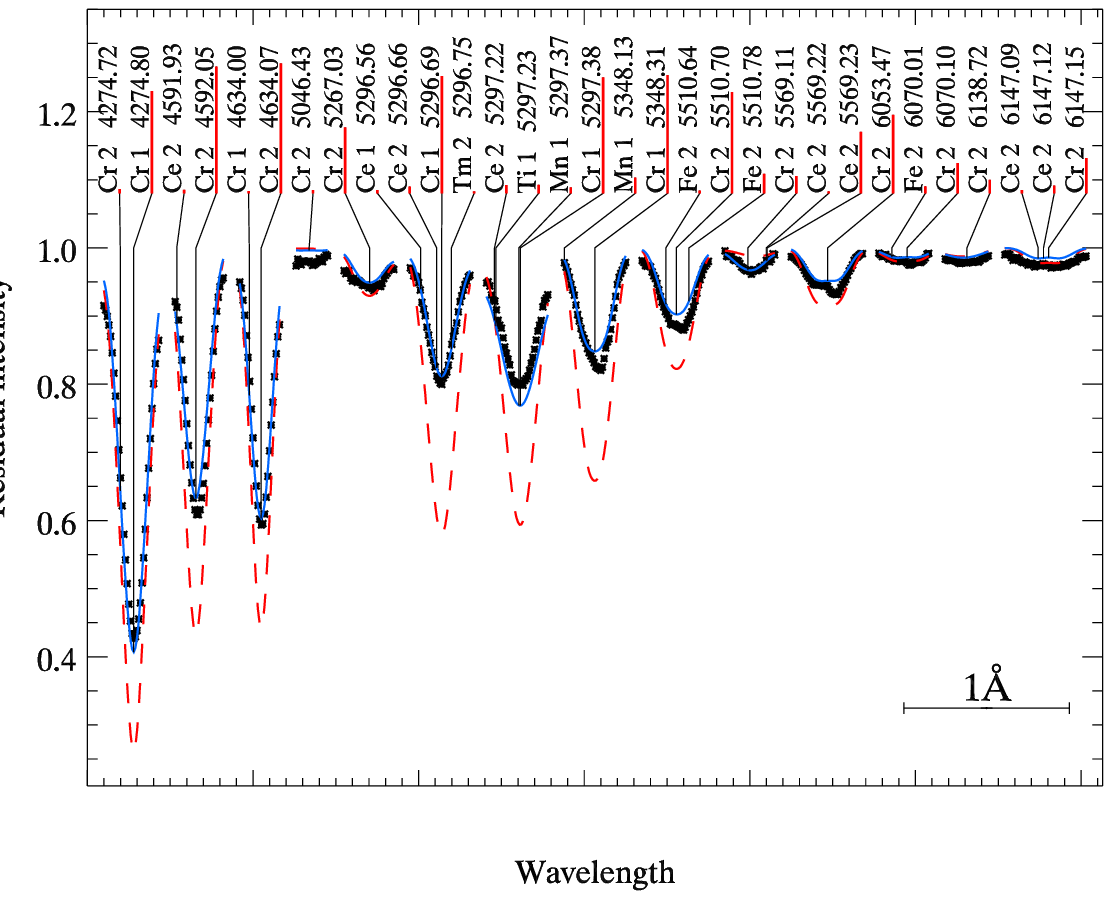}
\caption{A comparison between the observed \ion{Cr} line profiles and calculations with
the stratified abundance distribution (full line) and with the homogeneous ($\log(Cr/N_{\rm total})=-5.80$) 
abundances (dashed line).}
\label{Cr}
\end{figure*}

\begin{figure*}
\includegraphics[height=12cm]{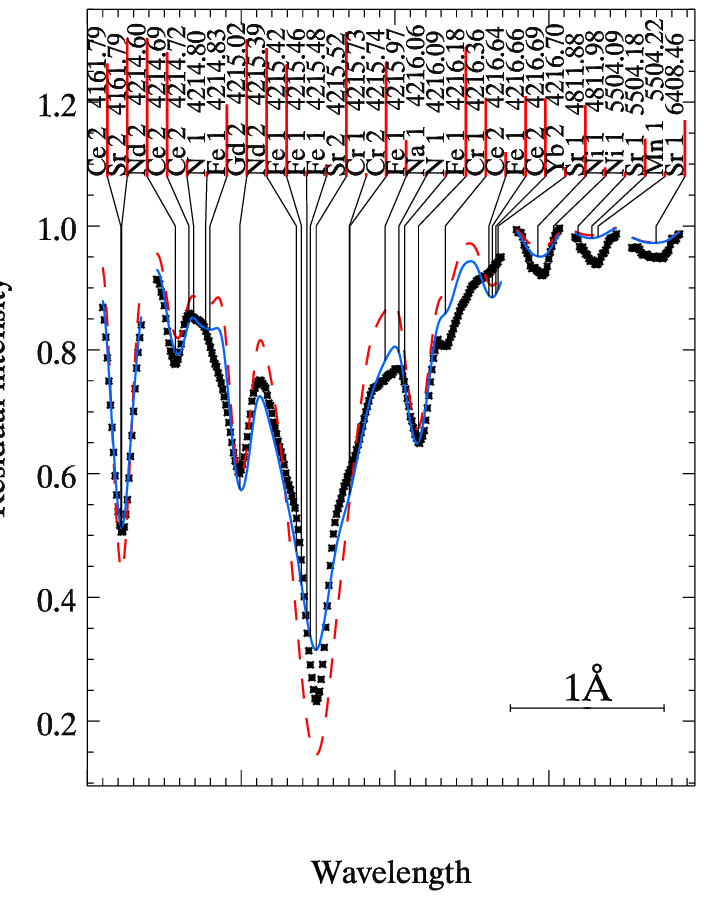}\hspace{2cm}\includegraphics[height=12cm]{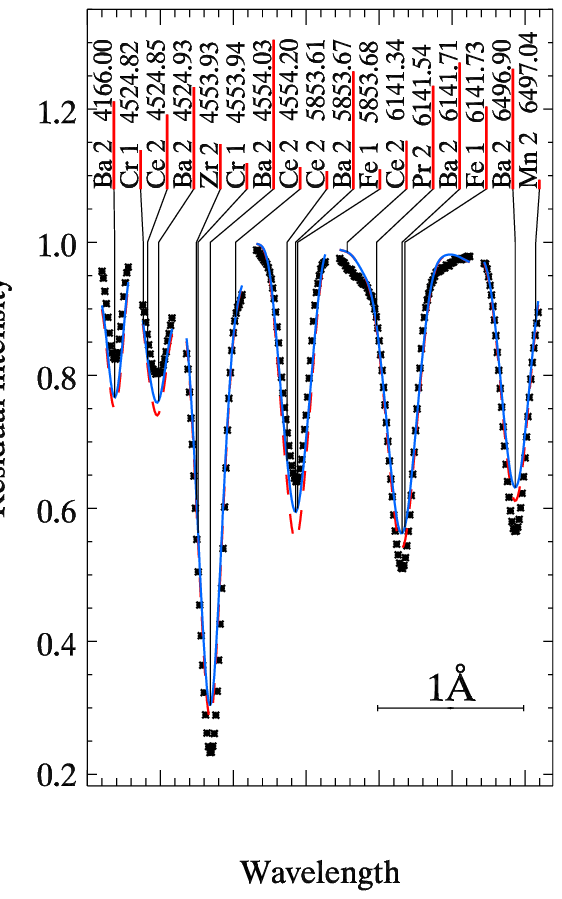}
\caption{A comparison between the observed \ion{Sr} (left panel) and \ion{Ba} (right panel) line profiles and calculations with
the stratified abundance distribution (full line) and with the homogeneous ($\log(Sr/N_{\rm total})=-8.00$, $\log(Ba/N_{\rm total})=-9.00$) 
abundances (dashed line).}
\label{Sr-Ba}
\end{figure*}

Fig.~\ref{Fsed} demonstrates the observed and predicted energy distributions of \hd.
Observational data is represented by a combination of the IUE low-resolution flux and photometric
measurements converted to absolute flux units.
The theoretical fluxes are calculated with the last model from Table~\ref{Tcolors}. Note that
all the models from Table~\ref{Tcolors} represent observations good enough, and the only possibility
to choose the best one would be the comparison of UV fluxes blueward $\lambda=2000$\AA\, where
stratified models predict significant flux excess (almost one order of magnitude in maximum) 
compared to homogeneous abundance model, but this region plays unimportant role in the total
energy balance in the atmosphere due to the relatively cool effective temperature of the star.
Although the IUE observations tend to be closer to the theoretical stratified fluxes 
at this region, it is difficult to make a clear decision because of large error bars in
observations (see IUE data archive).

Fitting of the spectral energy distribution of \hd\ calibrated in physical flux units allows us to derive
the angular diameter of the star, $\theta=0.335\pm0.005$~mas. Combined with the revised Hipparcos parallax
of $\pi=20.32\pm0.39$~mas \citep{par}, this yields a new determination of the stellar radius, 
$R=1.772\pm0.043R_{\odot}$.

\subsection{Effect of REE stratification}\label{effectree}
It was shown above that the accumulation of Pr and Nd leads to the appearance of inverse 
temperature gradient with
the temperature jump of about $600$\,K as compared to non-stratified model. 
However, this temperature jump is located in
upper atmospheric layers and has no influence on the wings of \halpha\, line, as it is shown in Fig.~\ref{Falpha}. 
Thus, even in the case of strong stratification of REE elements, the determination of atmospheric stellar parameters by
fitting the wings of hydrogen lines could be performed with the model not taking the stratification of REE into account.

However, LTE calculations show that the core of \halpha\, line is influenced strongly by stratification of Pr and Nd
(bottom panel of Fig.~\ref{Falpha}). One could see that none of the \halpha\, profiles computed under the 
assumption of LTE line formation could fit the transition region between the line core and wing,
that is commonly called as core-to-wing anomaly (CWA) and is frequently observed in spectra of Ap stars. 
\citet{coretowing} tried to explain this anomaly by the existence of the temperature bump 
in the upper atmospheric layers above $\log\tau_{\rm 5000}\approx-1$.
What is interesting is that the heating of atmospheric layers due to REE stratification 
produces the same sharp transition between the line wings and core,
but the central intensity of theoretical spectra are much shallower than that 
with homogeneous abundance model which still
provides the best fit. However, all stratification effects disappear when NLTE hydrogen lines formation
is applied, which is 'washing out' all stratification effects.

From Figs.~2 and~3 of \citet{coretowing} it is seen that in order to explain 
the core-to-wing anomaly of hydrogen Balmer lines, the
temperature jump should be located between $\log\tau_{\rm 5000}\approx-3$ and $-2$. In our calculations the temperature 
gradient starts higher in stellar atmosphere at $\log\tau_{\rm 5000}\approx-3$. 
Position of the temperature gradient caused by REE stratification depends on the details of NLTE line formation, but it is
practically impossible to shift REE jump position in \hd\, downward by more than 0.3 -- 0.5 dex in  
$\log\tau_{\rm 5000}$ scale \citep{pr-NLTE}.

REE stratification is not the only reason for inverse temperature gradients in Ap star atmosphere. Recently, \citet{diff2}
and \citet{LeBlanc09} calculated element stratification in the atmospheres with magnetic field and showed that horizontal
magnetic field changes abundance distribution. Instead of a step-like shape which is obtained in diffusion calculations in non-magnetic
atmospheres or in the presence of predominantly vertical component of the field, one gets an abundance
minimum and then an increase of element abundance in the uppermost atmospheric layers. 
These abundance increase might lead to the corresponding temperature increase started just at $\log\tau_{\rm 5000}=-1$
\citep[see Fig.5 of][]{LeBlanc09}. In a simple step function approximation used in the present stratification analysis as well as in the most other
similar studies \citep{Wade01,str2,str1,str4}, we cannot probe an increase of element abundance in the
uppermost atmospheric layers.
However, relatively good fitting of the observed line profiles in many stars with different strengths of magnetic field 
obtained with step-like abundance distribution of Si, Ca, Cr, Fe -- main contributors to line opacities -- supports 
the absence of significant increase of 
element abundance immediately following the abundance jump. This is further corroborated by the vertical inverse problem solution 
for the Ap star HD\,133792 \citep{vip}, where the authors were able to derive
element distribution without {\it a priori} assumption about the shape of this distribution. One should note,
however, that this method can be currently applied only to non-magnetic or weakly magnetic
atmospheres. Additionally, in atmospheric diffusion calculations the stratification is generally obtained
assuming that all the forces acting on a given
plasma volume are in stationary equilibrium (total diffusion velocity is zero). Regarding to real stars this 
is not necessarily the case and element stratifications at a given moment could be different 
\cite[see][]{diff2}.
       
Thus, to explain the most part of the observed peculiarities in cool Ap stars we need to implement the full NLTE treatment of REE elements 
in model atmosphere calculations and not the rough simulation presented in this study. 
The accumulation of other elements could also be responsible 
for CWA and more detailed investigation is needed to explore this effect.

\section{Conclusions}\label{ciao}
The spectroscopic investigation of elements stratification is the only method for testing modern
predictions of the particle diffusion theory in atmospheres of star where this mechanism
is sufficient to produce detectable abundance gradients. 
It provides us with stratification profiles of
different chemical elements that could be compared with recent self-consistent diffusion models. However,
in most of the stratification analysis routines the standard stellar model atmospheres are used where the 
homogeneous abundances are assumed.

To circumvent this difficulty we made an attempt to construct an empirical self-consistent model atmosphere 
where the stratification of chemical elements is derived directly from observed spectra and then treated
in a model atmosphere code. This iterative procedure was applied to one of the cool CP star \hd. 
Below we summarise the main results of this investigation:

\begin{itemize}
\item
The accumulation of Pr and Nd in the upper atmosphere has an 
impact 
on the temperature structure of the atmosphere.
The strong overabundance of these elements leads to an appearance of inverse temperature 
gradient with the maximum temperature increase of up to
$600-800$\,K as compared to homogeneous abundance model.
\item
For \hd\, we find that the effect of elements stratification has minor influence on the observed energy distribution.
The lack of good quality observations in UV region (where the effect of stratification is more pronounced) 
does not allow at present to carry out more precise investigations.
The changes in temperature structure due to the stratification of Pr and Nd
do not affect the wings of hydrogen lines and thus can be ignored in routine procedures
of the determination of atmospheric stellar parameters.
\item
Due to their high overabundance and important role in overall temperature balance, 
REE element can no longer be considered as trace elements in standard scheme of NLTE calculations. 
This implies that, to achieve a better consistency
in modelling of REE stratification, the model atmosphere should account for REE NLTE effects as well.
\item
For \hd\, we find no critical changes in the shape and amplitude of stratification profiles of chemical
elements when recalculating the atmospheric temperature-pressure structure. Most of the changes are
comparable with the error bars of the minimization procedure used.
\end{itemize}

\begin{acknowledgements}
This work was supported by following funding projects: 
FWF Lisa Meitner grant Nr. M998-N16 (DS), FWF P17890-N2 (TR), RFBR 08-02-00469-a and
Presidium RAS Programme ``Origin and evolution of stars and galaxies'' (TR and LM).
We also acknowledge the use of electronic databases (VALD, SIMBAD, NASA's ADS).
\end{acknowledgements}

\Online
%

\end{document}